\documentclass{emulateapj}

\usepackage{epsf}
\usepackage{graphics}
\usepackage{color}

\newcommand{\be}{\begin{equation}}
\newcommand{\ee}{\end{equation}}

\newcommand{\ax}{$\alpha_{\rm X}$}
\newcommand{\bx}{$\beta_{\rm X}$}

\newcommand{\rb}[1]{\raisebox{1.5ex}[-1.5ex]{#1}}

\newcommand{\plm}{$\pm$}

\newcommand{\swift}{{\it Swift}}

\newcommand{\chandra}{{\it Chandra}}
\newcommand{\meszaros}{M\'esz\'aros~}

\shorttitle{Chandra observation of GRB 060729}
\shortauthors{Grupe et al.}

\begin{document}

\def\etal{{\it et\thinspace al.}\ }
\def\alp{{$\alpha$}\ }
\def\al2{{$\alpha^2$}\ }

%
%
%


\title{Late-time detections of the X-ray afterglow of GRB 060729 with Chandra -
the latest detections ever of an X-ray afterglow
}


\author{Dirk Grupe\altaffilmark{1}
\email{grupe@astro.psu.edu},
David N. Burrows\altaffilmark{1},
Xue-Feng Wu\altaffilmark{1,2},
Xiang-Yu Wang\altaffilmark{3,1},
Bing Zhang\altaffilmark{4},
En-Wei Liang\altaffilmark{5},
Gordon Garmire\altaffilmark{1},
John A. Nousek\altaffilmark{1},
Neil Gehrels\altaffilmark{6}
George R. Ricker\altaffilmark{7},
Marshall W. Bautz\altaffilmark{7}
}

\altaffiltext{1}{Department of Astronomy and Astrophysics, Pennsylvania State
University, 525 Davey Lab, University Park, PA 16802}

\altaffiltext{2}{Purple Mountain Observatory, Chinese Academy of Sciences, Nanjing 210008, China}

\altaffiltext{3}{Department of Astronomy, Nanjing University,
Nanjing 210093, China}

\altaffiltext{4}{Department of Physics, University of Nevada, Las Vegas,
NV 89154}

\altaffiltext{5}{Department of Physics, Guangxi University, Nanning 530004, China}

\altaffiltext{6}{Astrophysics Science Division, Astroparticle Physics Laboratory,
Code 661, NASA Goddard Space Flight Center, Greenbelt, MD 20771 }

\altaffiltext{7}{Massachusetts Institute of Technology,
77 Massachusetts Av., Cambridge, Ma 02139-4307}




\begin{abstract}
We report on 5 \chandra\ observations of the X-ray afterglow of
the Gamma-Ray Burst GRB 060729 performed between 2007 March and
2008 May. In all five observations the afterglow is clearly
detected. The last  \chandra\ pointing was performed on
2008-May-04, 642 days after the burst - the latest detection of a
GRB X-ray afterglow ever. A reanalysis of the \swift\ XRT light
curve together with the three detections by \chandra\ in 2007
reveals a break at $\sim$1.0 Ms after the burst with a slight
steepening of
 the decay slope from $\alpha$ = 1.32 to 1.61. This break
coincides with a significant hardening of the X-ray spectrum, consistent with a cooling break in the wind medium
scenario, in which the cooling frequency of the afterglow 
crosses the X-ray band.
The last two \chandra\ observations in 2007 December
and 2008 May provide evidence for another break at about one year
after the burst. If interpreted as a jet break, this late-time
break implies a jet half opening angle of $\sim14^{\circ}$ for a
wind medium.
Alternatively, this final break may have a spectral origin, in which case no jet break
 has been observed and the half-opening
angle of the jet of GRB 060729 must be larger than $\sim
15^{\circ}$ for a wind medium. We compare the X-ray afterglow of
GRB 060729 in a wind environment with other bright X-ray
afterglows, in particular GRBs 061121 and 080319B, and discuss why
the X-ray afterglow of GRB 060729 is such an exceptionally
long-lasting event.
\end{abstract}


\section{Introduction}

Gamma-Ray Bursts (GRBs) are the most energetic transient events in
the Universe. The most accepted model to explain this phenomenon
is the ``fireball" model \citep[e.g. ][]{meszaros06, zhang04} and
the progenitor of long GRBs are believed to be massive stars on
the order of 30 solar
 masses or more.
 The isotropic energies
inferred from the observed fluences are often of order $10^{53} -
10^{54}$ ergs. On the other hand, typical supernova explosions are
of the order of only $10^{51}$ ergs. One way to solve this
``energy problem" is by assuming that the radiation is collimated
into a jet. One prediction of the ``fireball" model is that the
afterglow decay rate increases when the relativistic beaming angle
equals or exceeds the physical jet opening angle as the jet
decelerates in the surrounding medium \citep[e.g.
][]{rhoads99,sari99}.
 This can be seen as an achromatic jet break in the light curve, with a typical
decay slope after the jet break of order $\alpha$ = 2 (e.g. Zhang
et al. 2006; Nousek et al. 2006). The time of this jet break can
be used to infer the jet opening angle by e.g. the relation given
by \citet{frail01}. The measurement of the jet break time is
therefore most critical for understanding the energetics of GRBs.

Before the launch of the {\it Swift}
Gamma-Ray Burst Explorer mission  \citep{gehrels04}, putative jet breaks were found at
optical or radio wavelengths, typically
 a few days after the burst
 \citep[e.g. ][]{frail01}. Since its launch, {\it Swift}
has detected roughly 400 bursts (end of 2008)
 and typically observes them
for up to one or two weeks after the trigger. For the majority of
these bursts, jet breaks have not been detected \citep[e.g.
][]{burrows07,willingale07, liang08, racusin09}. However, one
reason could be that jet breaks occur in X-rays at much later
times than previously thought and the follow-up observations by
{\it Swift} are not late enough in time to detect a jet break, as
studies by \citet{willingale07} and \citet{sato07} suggest. As an
alternative, \citet{curran08} suggested that the jet breaks are
hidden and the light curves are mis-interpreted as a single power
law decay although a jet break is there. Only a handful of
afterglows have been followed for more than a month by \swift,
because usually the X-ray afterglow fades below the \swift\ 
 X-Ray telescope \citep[XRT; ][]{burrows05} detection
limit  $\sim 10^{-14}$ ergs s$^{-1}$ cm$^{-2}$ roughly
 a week after the trigger. One of these exceptions is
the bright X-ray afterglow of GRB 060729 which was detected by
\swift\ XRT
 even 125 days after the burst
\citep[][]{grupe07}.

The \swift~Burst Alert Telescope \citep[BAT; ][]{barthelmy05}
triggered on GRB 060729 on  2006 July 29, 19:12:29 UT
\citep{grupe07} and a redshift of $z=0.54$ was measured
\citep{thoene06}.
 The
XRT and the UV/Optical Telescope \citep[UVOT; ][]{roming05},
started observing the burst about 2 minutes after the trigger. The
\swift\ UVOT was able to follow this afterglow in the UVW1 filter
up to 31 days after the BAT trigger. In X-rays \swift's XRT was
still detecting the X-ray afterglow of GRB 060729 at the end of
2006 November,  125 days after the burst \citep{grupe07}. However,
by 2006 December the \swift-XRT detection limit  was reached and
only a 3$\sigma$ upper limit could be given for the 63.5 ks
exposure time obtained in December 2006. By that time the X-ray
afterglow of GRB 060729 did not show any clear evidence for a jet
break, giving a lower limit on the jet opening angle of
$\theta=28^{\circ}$ \citep{grupe07} based on the assumption of a
constant circumburst medium. In order to extend the light curve of
this exceptional X-ray afterglow, we  observed it five times with
\chandra\ ACIS in 2007 and 2008.

We report on the detections of the X-ray afterglow of GRB 060729
\citep{grupe07} with \chandra\ up to 642 days after the burst -
the latest detection ever of an X-ray afterglow of a GRB at
cosmological distance. Previously the burst with the latest
detection of an X-ray afterglow was GRB 030329 \citep{tiengo03,
tiengo04}, which had a detection 258 days after the burst by {\it
XMM-Newton}. Our paper is organized as follows: in
\S\,\ref{observe} the observations and data reduction are
explained, in \S\,\ref{results} the measurements of the X-ray
light curve are shown, and in \S\,\ref{discuss} we discuss the
implications of this light curve. Throughout this paper the X-ray
flux dependence on time and frequency is defined as $F\propto
t^{-\alpha}\nu^{-\beta}$. Luminosities are calculated assuming a
$\Lambda$CDM cosmology with $\Omega_{\rm M}$ = 0.30,
$\Omega_{\Lambda}$ = 0.70 and a Hubble constant of $H_0$ = 71 km
s$^{-1}$ Mpc$^{-1}$ corresponding to
 a luminosity distance $d_L$ = 3064 Mpc for GRB 060729.

\section{Observations and data reduction}
\label{observe}

\chandra\ observed GRB 060729
three times between 2007 March 16 and 2007 June 30.
Two very late-time \chandra\  observations were performed in 2007 December/2008 January for 72.7 ks
and in 2008 April/May for 117.3 ks.
Due to pitch angle constraints some of these had to be split into several visits.
 All observations, with start and end times and exposure times, are listed
in Table\,\ref{xray_log}.

All of these observations were performed with the standard 3.2s
readout time in Very Faint mode on the on-axis position on the
back-illuminated ACIS-S3 CCD. Data reduction was performed with
the \chandra\ analysis software CIAO version 4.0 and the
calibration database CALDB version 3.4.3. In order to reduce the
ACIS particle background all \chandra\ stage 1 event data were
reprocessed using CIAO {\tt acis\_process\_events} with the VF
mode cleaning. Only ACIS grades 0, 2, 3, 4, and 6 were selected
for further analysis. The background was further reduced by using
only photons in the 0.5 - 8.0 keV energy range. Before further
analysis, the observations were combined into one event file each
using the CIAO task {\it merge\_all}. Source photons were selected
in a circle with a radius $r$ = 1$^{''}$ and background photons in
a close-by source-free region with a radius $r$ = 10$^{''}$. Count
rates were converted into fluxes by using PIMMS version 3.9b using
the parameters from the spectral fits to the \swift\ data after
the break at 1 Ms after the burst (see below) with an absorption
column density $N_{\rm H}=1.34 \times 10^{21}$ cm$^{-2}$ and an
X-ray spectral index \bx = 0.89.

A description of the reduction and analysis of the \swift\ data
can be found in \citet{grupe07}. For display purposes and fitting
the late-time light curve we rebinned the \swift\ XRT Photon
Counting data with 250 counts per bin for the times up to 2 Ms
after the burst and 100 counts per bin for the times thereafter.
Spectral analyses were performed for the times 300 - 800 ks and $T
>$ 1 Ms after the burst. Source photons were collected in a circle
with $r = 23\farcs5$ and background photons with $r = 96^{''}$
with grade selection 0 - 12. The response matrix {\it
swxpc0to12s0\_20010101v010.rmf} was used. The spectra were
rebinned with 20 counts per bin for the 300-800 ks after the burst
spectrum and 15 counts per bin for the spectrum with $T >$ 1 Ms.
The spectra were  analyzed with XSPEC version 12.4.0x
\citep{arnaud96}. To search for changes in the X-ray spectrum we
applied a hardness ratio study\footnote{{\label{f7}}The hardness
ratio is defined as HR=(H-S)/(H+S), where S and H are the observed
counts in the 0.3-2.0 and 2.0-10.0 keV energy bands,
respectively.} segment by segment. Because of the low-number
statistics in some of the later segments of the \swift\ XRT and
all \chandra\ observations, we applied Bayesian statistics to
determine the hardness ratios as described by \citet{park06}.

\section{Results}
\label{results}

\subsection{Temporal breaks}

\begin{figure}
\epsscale{0.7} \plotone{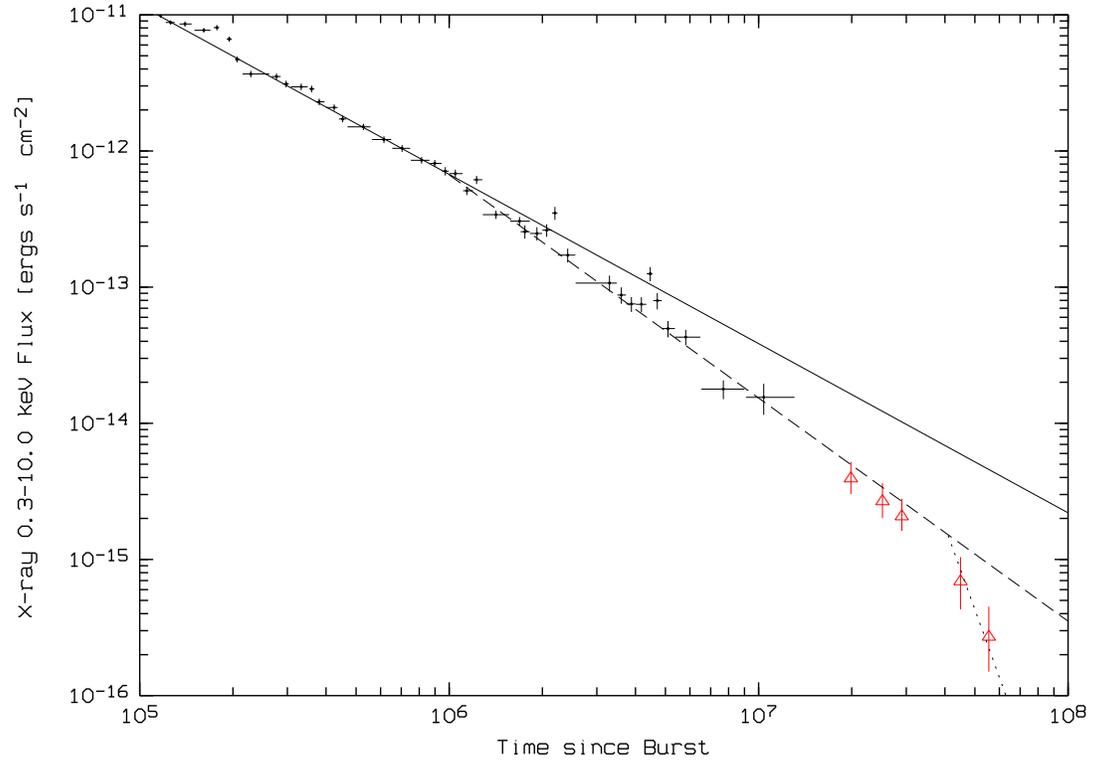} \caption{\label{xray_lc} \swift-XRT
and \chandra\ ACIS-S light curve of the X-ray afterglow of GRB
060729. The black crosses display the \swift\ XRT data and the red
triangles the \chandra\ data. The solid line displays the initial
decay slope $\alpha_3$=1.32 as reported by \citet{grupe07}, the
dashed line the decay slope post-break at 1.0 Ms after the burst
with $\alpha_4$=1.61 (from model 6 in Table 2), and the dotted line the steep decay after
the jet break at 41 Ms after the burst with $\alpha_5=4.65$. }
\end{figure}

The late-time
X-ray light curve including the five \chandra\ pointings is shown in
Figure\,\ref{xray_lc}.
We ignore the first day of the \swift\ observation because it is not relevant for the study of the
late-time light curve. The early light curve and a detailed discussion can be
found in \citet{grupe07}.

The late-time light curve (Figure\,\ref{xray_lc}) was fitted by
several power law and multiply-broken power law models as listed
in Table\,\ref{lc_fits}. Fitting the light curve with the decay
slope $\alpha_3$\footnote{We followed the notation as defined in
\citet{nousek06} and \citet{zhang06} where $\alpha_3$ is the steep
decay slope after the plateau phase.} fixed to 1.32, as reported
by \citet{grupe07} from the \swift\ data, gives a very poor result
($\chi^2/\nu$ = 897/46).
 The fit can be improved by leaving the decay slope as a free parameter (Table\,\ref{lc_fits}, model 2).
This results in a single decay slope of $\alpha_3$ = 1.45\plm0.01,
but the light curve still deviates significantly at later times
from this slope, resulting in an unacceptable $\chi^2/\nu$ =
400/45. A broken power law fit to the entire late-time light curve
(model 3) reveals a break at about 2 ~ Ms after the burst; in
contrast to the result of \citet{grupe07}, in which we could fit
the late-time \swift\ data with just one decay slope, the addition
of the 2007 \chandra\  data requires a break in the \swift\ data.
The late decay slope, $\alpha_4$ = 1.85, is driven by the last two
\chandra\ observations, while the $\chi^2$ is also strongly
affected by two very high data points at $\sim2$ Ms and $\sim5$
Ms. Making the assumption that these two high points are late-time
X-ray flares unrelated to the afterglow of the external shock, we
removed them from further fits (see models 4 and 5).
We then fit the data between
100~ks and 30~Ms with a broken power law (model 6), obtaining a
break time of $T_{\rm break,3} = 1.01^{+0.35}_{-0.22}$ Ms and
slopes of
 $\alpha_3 = 1.32^{+0.02}_{-0.05}$ and  $\alpha_4 = 1.61^{+0.10}_{-0.06}$.
This fit is plotted as the dashed line in Figure\,\ref{xray_lc}.

The last two \chandra\ observations deviate  from this fit,
suggesting a  break at about a year after the burst. Because these
last two points have very few counts (and consequently large
uncertainties), a broken power law fit to the late-time light
curve cannot constrain either the break time or the late-time
decay slope $\alpha_5$ unless at least one parameter is fixed. We
therefore approached the question of a final break in steps. A
single power law fit to the light curve for $T \geq 1.2$ Ms
(Table\,\ref{lc_fits}, model 7) gives $\alpha_4$ = 1.68\plm0.08
and $\chi^2/\nu$ = 12/15. Although this is already an acceptable fit,
we investigated the possibility of a late-time break which is expected
from GRB theory \citep[c.f. e.g.][]{zhang06, meszaros06}.
At first we fitted  the light
curve for $1.2$ Ms $\leq T \leq$ 35 Ms with a single power law
(model 8) which results in $\alpha_4$
= $1.61^{+0.07}_{-0.13}$ and $\chi^2/\nu$ = 6/13. We then fitted a
broken power law model to the entire light curve with $T
>$ 1.2 Ms with  $\alpha_4$ fixed at 1.61 (the best-fit result when
the last two \chandra\ observations are excluded; model 8 in
Table\,\ref{lc_fits}) to determine whether the data require a very
late break in the light curve slope. This fit
(Table\,\ref{lc_fits}, model 9) gives $T_{\rm break,4}$ =
41.3$^{+4.2}_{-5.1}$ Ms and $\alpha_5=4.65^{+2.05}_{-1.34}$, with
$\chi^2/\nu=6/14$.  Figure\,\ref{contour} 
displays the contour plot
between the final break time and the final slope. It shows that
they are still not well-constrained. Although the best-fit break
time is 41 Ms (2007 November), a break as early as $\sim 26$ 
Ms, with a late-time decay slope of $\alpha_5$ = 2.5,
is consistent with the data at the $1 \sigma$ level.

In addition to the broken power law fits with a sharp break, the
late-time light curve was also fitted by the smoothed
double-broken power law model defined by \citet{beuermann99}. 
Here we found decay slopes $\alpha_3$=1.20\plm0.07,
$\alpha_4$=1.70\plm0.05, and $\alpha_5$=2.41\plm0.26 with break
times at 1.09\plm1.01 Ms and 20 Ms (fixed). The smooth parameter
is fixed to 3.0 and 2.0 for the breaks at about 1 Ms and 20 Ms,
respectively. This results in an acceptable fit with $\chi^2/\nu$
= 42/32. 
Possible interpretations of these temporal breaks are discussed in \S\ref{discuss}.

\subsection{Spectral variations}

\begin{figure}
\epsscale{0.7} \plotone{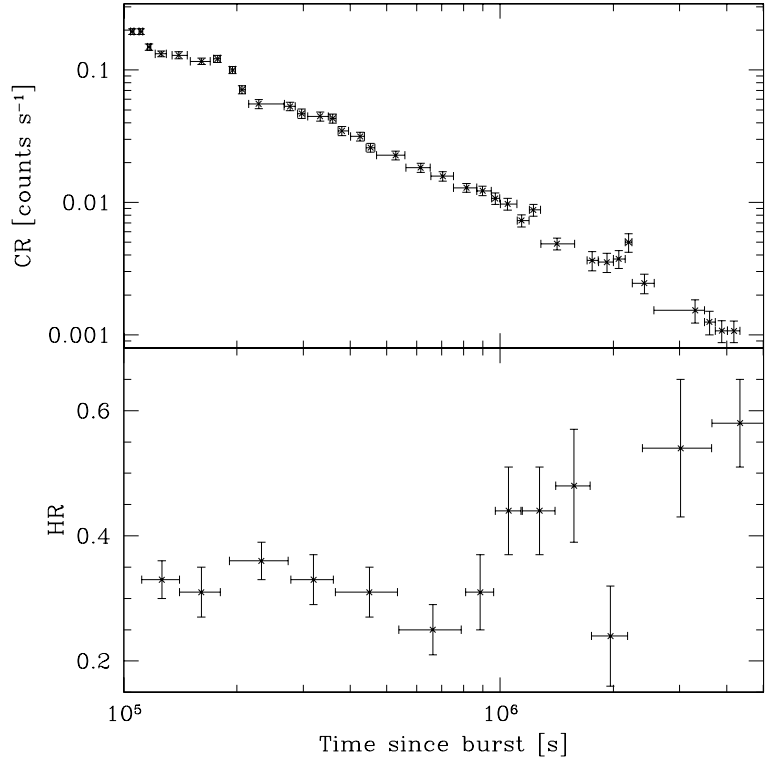} \caption{\label{lc_cr_hr}
\swift-XRT count rate and hardness ratio (as defined in section 3.1, footnote 8)
light curves
for the time around the break at 1.0 Ms after the burst.}
\end{figure}

\begin{figure}
\epsscale{0.7} \plotone{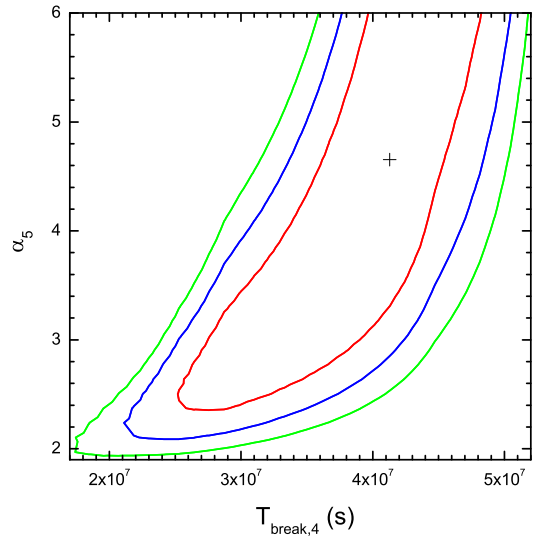}
\caption{\label{contour} Contour plot between the late-time break
and decay slope $\alpha_5$ of the fit to the late time light curve
of GRB 060729 with $T>$1.2 Ms after the burst, excluding the flares at 2 and 5 Ms (Model \#9 in Table\,\ref{lc_fits}).
The lines mark the
1, 2, and 3 $\sigma$ confidence levels. }
\end{figure}

Temporal breaks are often associated with spectral breaks
\citep[e.g. ][]{sari98, meszaros98, zhang06}.
Figure\,\ref{lc_cr_hr} displays the \swift\ XRT count rate and
hardness ratio light curves for the interval between 100 ks and 5
Ms after the burst. The hardness ratios are plotted segment by
segment. While the hardness ratios before the break at 1 Ms after
the burst are of  order  HR $\sim 0.3$, after the break the
spectrum hardens to HR $\sim 0.45$ with even harder values at
later times.

The spectrum before the 1 Ms break can be fitted with a single
absorbed power law with $N_{\rm H} = (1.34^{+0.27}_{-0.25})\times
10^{21}$ cm$^{-2}$ and an energy spectral slope $\beta_{\rm x}$ =
1.18\plm0.11 ($\chi^2/\nu$ = 81/82). The spectrum after the 1 Ms
break was also fitted by an absorbed single power law model.
Leaving the absorption column density as a free parameter,
however, results in an increase of the column density, which does
not seem plausible. Therefore we fixed the absorption column
density to $N_{\rm H} = 1.34\times 10^{21}$ cm$^{-2}$, the value
obtained before the break. This fit results in a slightly flatter
energy spectral slope $\beta_{\rm x}$ = 0.89\plm0.11. These values
were used in PIMMS to convert the \chandra\ ACIS-S count rates
into the fluxes given in Table\,\ref{xray_log} and plotted in
Figure\,\ref{xray_lc}.

The \chandra\ data must be analyzed in the Poisson limit,
complicating proper analysis of possible spectral variations at very late times.
Using the Bayesian approach described by \citet{park06}, we
estimated the hardness ratios in the \chandra\ data\footnote{We defined the
\chandra\ hardness ratio as HR=(H-S)/(H+S), where S and
H are the counts in the 0.5-2.0 and 2.0-8.0 keV bands,
respectively.} and their uncertainties, both before and
after the break at 38 Ms.
We obtain mean values of HR=$-0.39$ for the 2007 March-June data (before the final break; 38 counts total) and HR=$-0.80$ for
the very late data (after the final break; 8 counts total), with 85\% confidence limits of HR$= -0.60$ to $-0.17$
and HR$= -1.00$ to $-0.58$, respectively.  Although this is a suggestion of spectral softening
across the final break, we cannot exclude (at the 85\% confidence level) the possibility that the hardness ratio
is constant.
Note that due to the different
energy bands and detector response matrices it is not possible to
compare the \swift\ and \chandra\ hardness ratios directly.

\subsection{Comparison with other GRBs}

\begin{figure}
\epsscale{0.6}
\plotone{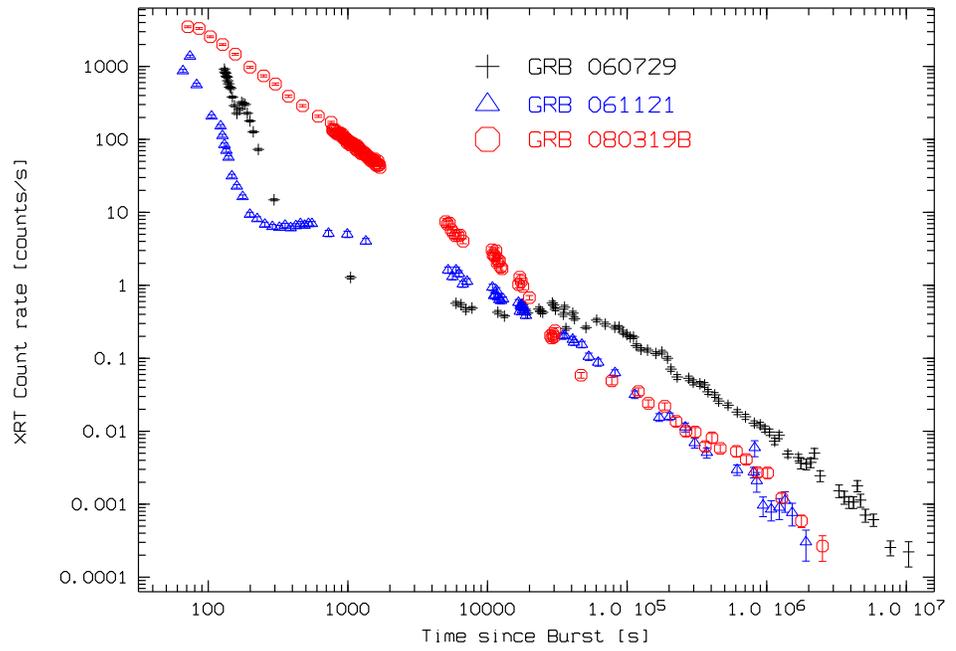}
\caption{\label{grbs060729_061121} Comparison of the observed X-ray light curves
of GRBs 060729, 061121, and GRB 080319B
}
\end{figure}

Even though GRB 060729 was one of the brightest bursts detected in
X-rays, it is not the brightest one so far seen during the
\swift\ mission. GRB 061121 was about 2-3 times brighter when the
\swift-XRT started observing it \citep{page07},
but by a day after the burst it was an order of magnitude fainter than
GRB 060729 and was not detected after 2 Ms post-burst.
 The second brightest X-ray afterglow so
far seen by  \swift\ was GRB 080319B (Racusin et al., 2008).
Figure \ref{grbs060729_061121} displays the observed count rate
light curves of GRBs 060729, 061121, and 080319B. Even though GRBs
061121 and 080319B appear to be much brighter than GRB 060729
until about 20 ks after the trigger, the long plateau phase in GRB
060729 makes it the brightest X-ray afterglow about half a day
after the trigger. GRB 061121 already displays a break from the
plateau to the second steep decay phase at 2.2 ks after the burst
followed by an even steeper decay at 30 ks with a decay slope of
$\alpha$ = 1.5 (Page et al., 2007). This earlier break compared to
GRB 060729, which broke at about 60 ks after the burst, and the
steeper decay slope of 1.5 compared to 1.3 in GRB 060729 made GRB
061121 disappear much faster than GRB 060729. The ``naked-eye"
burst 080319B on the other hand does not even show a noticeable
plateau phase and decays rather quickly with late-time decay
slopes of \ax = 1.17 and 2.61 before and after the late-time jet
break at $T=9.5\times10^{5}$ s.

\begin{figure}
\epsscale{0.6}
\plotone{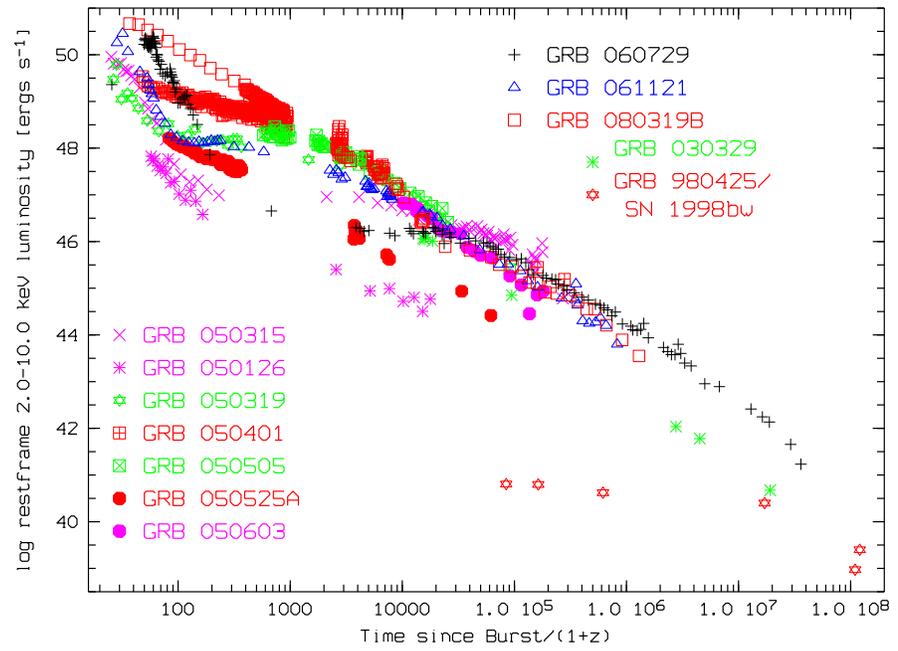}
\caption{\label{grbs_lc_l_t} Rest frame isotropic equivalent luminosity light curves of several
\swift\ and pre-\swift\ bursts taken from \citet{nousek06} including GRBs 060729,
061121, and 080319B.
}
\end{figure}

We can ask how the intrinsic rest-frame 2-10 keV luminosity light
curve of GRB 060729 compares to the bursts shown in Figure 1 in
\citet{nousek06}. Figure\,\ref{grbs_lc_l_t} displays GRBs 060729,
061121, and 080319B with most of the bursts shown in Figure\,1 in
\citet{nousek06} in the rest-frame. The plot shows that during the
plateau phase GRB 060729 was not that luminous. As a matter of
fact it was a factor of about 10 to 100 less luminous than bursts
such as GRBs 061121 or 080319B. Nevertheless after a few days in
the rest-frame, GRB 060729 becomes the most luminous X-ray
afterglow in the 2 - 10 keV band.

Compared with other GRBs shown in Figure\,\ref{grbs_lc_l_t}, the
total energy output in the apparent rest-frame 2-10 keV band of
$E_{\rm 2-10 keV} = 7\times 10^{52}$ ergs makes GRB 060729 one of
the most energetic X-ray afterglows ever detected. In the 2-10 keV
band, only GRBs 061121 and 080319B appear to be more energetic
with $E_{\rm 2-10 keV} = 1\times 10^{53}$ ergs and $2\times
10^{54}$ ergs, respectively. However, if we attempt to correct for
the jet opening angle, the picture changes. With jet opening
angles of 4$^{\circ}$ and 0.4$^{\circ}$ as inferred for GRBs
061121 and 080319B, respectively \citep{page07, racusin08}, the
beaming-corrected energies are $2.4\times 10^{50}$ and $4.0\times
10^{50}$ ergs, respectively. For GRB 060729, however, assuming a
jet
 half opening angle of 14$^{\circ}$ (see next section) the
beaming corrected energy is still $2.1\times 10^{51}$ ergs in the
rest frame 2 - 10 keV band.

\section{Discussion}
\label{discuss}

While our original \swift\ XRT light curve after the plateau phase
seems to be consistent with a single power law decay
\citep{grupe07}, the \chandra\ data make it apparent that there
had to be a break at about 1 Ms after the burst. We showed above
that this break coincides with a significant change in the X-ray
spectral slope from \bx = 1.2 before and 0.9 after that break.
Unfortunately, as mentioned in \citet{grupe07}, \swift\ could only
follow the afterglow in the UVOT with the W1 filter up to a month
after the burst due to bright stars in the UVOT field which caused
some scatter in the UVOT and an enhanced background at the
position of GRB 060729.

According to Model 9 (excluding the two X-ray flare points at 2 Ms
and 5 Ms) in Table 2, the late time X-ray afterglow is described
by $F_{\nu_X}\propto t^{-\alpha_i}\nu_{X}^{-\beta_i}$, where (1)
$\alpha_3=1.32^{+0.02}_{-0.05}$, $\beta_3=1.18\pm0.11$ for 105 ks
$<t< T_{\rm break,3}=1.01^{+0.35}_{-0.22}$ Ms; (2)
$\alpha_4=1.61^{+0.10}_{-0.06}$, $\beta_4=0.89\pm0.11$ for $T_{\rm
break,3} < t < T_{\rm break,4} = 4.13^{+0.42}_{-0.51} \times 10^7$
s; (3) $\alpha_5=4.65^{+2.05}_{-1.34}$ for $t
> T_{\rm break,4}$.

The closure relation for 0.1 Ms $< t < 1.0$ Ms results in
$\alpha_3-1.5\beta_3=-0.46 \pm 0.17$, consistent with both the ISM
and wind models with theoretical expectation of
$\alpha_3-1.5\beta_3=-0.5$ if $\nu_X
> >$ max$\{\nu_c,\nu_m\}$, so $\alpha_3 = (3p-2)/4$ and $\beta_3 = p/2$.
The power law index of the electron energy distribution is $p =
2.43^{+0.03}_{-0.07}$ (2.36\plm0.22), derived from
$\alpha_3$($\beta_3$). The closure relation for $1.0$ Ms $< t < 41$
Ms results in $\alpha_4-1.5\beta_4=0.28^{+0.19}_{-0.18}$,
moderately consistent with the theoretical expectation of
$\alpha_4-1.5\beta_4=0.5$ if the environment is a free
wind\footnote{For the ISM model, the closure relation is
$\alpha_{4}-1.5\beta_{4}=-0.5$ or 0.0, which can be excluded at
the 5$\sigma$ and 2$\sigma$ confidence level, respectively. Even
if we adopt $\alpha_{4}-1.5\beta_{4}=0$, the derived
$p\sim3.15^{+0.13}_{-0.08}/2.78\pm0.22$ from the temporal/spectral
index is inconsistent with the $p$ value derived with $\alpha_3$
and $\beta_3$ at the earlier stage.} and the spectral regime is
$\nu_m< \nu_X < \nu_c$, so $\alpha_4 = (3p-1)/4$ and $\beta_4 =
(p-1)/2$. In other words, the break at $\sim 1.0$ Ms can be
interpreted as a cooling break in the wind medium scenario in
which the cooling frequency $\nu_c$ 
crosses the X-ray band. The power law index of the electron energy
distribution derived from the value of $\alpha_4$ (the spectral
index of this epoch has a relatively large uncertainty) is $p =
2.48^{+0.13}_{-0.08}$, quite consistent with the value derived
during the previous epoch. Therefore, a wind model is preferred
from the observations before $t = 41$ Ms, breaking the degeneracy
of the wind and ISM models which are both consistent with the
earlier data \citep{grupe07}.

 The X-ray afterglow light curve for $t>2.75\times10^5$ s
excluding the two flares can be also well fitted by a single
smoothed broken power law: (1) $\alpha_3=1.31\pm0.05$,
$\beta_3=1.18\pm0.11$ for 275 ks $<t< T_{\rm break,3}=2.43\pm0.79$
Ms; (2) $\alpha_4=1.96\pm0.09$, $\beta_4=0.89\pm0.11$ for $T_{\rm
break,3} < t$ ($\chi^2_{\nu}/{\rm dof} = 42/32$, smoothness
parameter $s = 3$). The steepening of the decay at $t=2.4$ Ms, if
not due to the cooling frequency passing through the observing
band (which results $\alpha_4-\alpha_3=0.25$, inconsistent with
the observation), should  originate from the post-jet-break
evolution. The change of the temporal decay index,
$\alpha_4-\alpha_3=0.65\pm0.10$, is consistent with a
non-spreading jet break in a wind model with $\Delta\alpha=0.5$
(if the jet has significant sideways expansion, then the value of
$\alpha_4$ should be equal to $p \sim 2.4 - 2.5$). However, the
spectral hardening around the break time can not be well
interpreted in such a model. Furthermore, the transition of a jet
break in the wind medium usually takes two orders of magnitude in
time (Kumar \& Panaitescu 2000), which is inconsistent with the
observation of GRB 060729. In conclusion, the temporal break at $t
\sim 1 - 3$ Ms is probably not a jet break.

There are two possible interpretations for the last tentative
light curve break at $t = 41$ Ms, as follows: a jet break, or a
spectral break in a spherical model. We discuss these next, after
which we consider the implications of the long plateau phase.

\subsection{Jet model} The jet + wind model predicts $\alpha_5 = p
\sim 2.4 - 2.5$ for a sideways expanding jet or $\alpha_5 =
\alpha_4 + 0.5 \sim 2.1$ for a non-sideways expanding jet. The
value of the model-predicted temporal index after the jet break
thus can not be excluded at even the $1\sigma$ confidence level (see
Fig.\,\ref{contour}). Recently Zhang \& MacFadyen (2009) have
performed two dimensional simulations and calculations of GRB
afterglow hydrodynamics and emission. They showed that the
sideways expansion of GRB jets can be neglected during the
relativistic phase (for the sideways expansion of GRB jets see
also Kumar \& Granot 2003, Granot \& Kumar 2003) and
that the change in decay slope was larger than predicted
analytically.
Their results may further alleviate the above problem of the
relatively shallow theoretical slope compared with the observed
steep slope.
However, the spectral softening revealed
by the hardness ratio evolution around this break time
somewhat disfavors the jet break interpretation.

If we interpret this break as a jet break, then a half-opening
angle of the jet can be inferred \citep[e.g.,][]{chevalier00}.
Under the thin-shell approximation for the post-shocked fluid of a
relativistic blast wave, the conservation of energy (neglecting
the initial baryon loading in the fireball) reads
\begin{equation}
E_{k,\rm{iso}}=M_{\rm sw}\Gamma^2c^2=\rm{constant},
\end{equation}
where $\Gamma$ is the bulk Lorentz factor of the downstream fluid
just behind the shock front and $\Gamma_s=\sqrt{2}\Gamma$ is the
Lorentz factor of the shock. The swept-up circum-burst mass by the
blast wave is
\begin{equation}
M_{\rm sw}=\int_{0}^{R}4\pi r^2 n(r)m_p dr =
\frac{4\pi}{3-k}AR^{3-k} m_p,
\end{equation}
where the environmental density $n(r)=Ar^{-k}$, and $A=n_0$ for
the ISM case ($k=0$) and $A=3\times10^{35}$ cm$^{-1}A_{\ast}$ for
the stellar wind case ($k=2$).

The evolution of the shock radius $R$ measured in the observer's
frame is $dR=2\Gamma_s^2cdt/(1+z)$, therefore,
\begin{equation}
R=\left[\frac{(3-k)(4-k)E_{k,\rm{iso}}t}{(1+z)\pi Am_p
c}\right]^{1/(4-k)},
\end{equation}
and
\begin{equation}
\Gamma=\left[\frac{(3-k)E_{k,\rm{iso}}}{4\pi AR^{3-k}m_p
c^2}\right]^{1/2}.
\end{equation}
Inserting the inferred values of the physical parameters, we have
\begin{equation}
R=1.2\times10^{19}E_{k,\rm{iso},54}^{1/4}n_{0,-1}^{-1/4}\left(\frac{t}{41\;\rm{Ms}}\right)^{1/4}\;\rm{cm},
\end{equation}
\begin{equation}
\Gamma=1.0E_{k,\rm{iso},54}^{1/8}n_{0,-1}^{-1/8}\left(\frac{t}{41\;\rm{Ms}}\right)^{-3/8},
\end{equation}
for the ISM case, and
\begin{equation}
R=1.0\times10^{20}E_{k,\rm{iso},54}^{1/2}A_{\ast,-1}^{-1/2}\left(\frac{t}{41\;\rm{Ms}}\right)^{1/2}\;\rm{cm},
\end{equation}
\begin{equation}
\Gamma=4.0E_{k,\rm{iso},54}^{1/4}A_{\ast,-1}^{-1/4}\left(\frac{t}{41\;\rm{Ms}}\right)^{-1/4},
\end{equation}
\begin{equation}
\theta_{\rm{jet}}=14.0^{\circ}E_{k,\rm{iso},54}^{-1/4}A_{\ast,-1}^{1/4}\left(\frac{t}{41\;\rm{Ms}}\right)^{1/4},
\end{equation}
for the wind case. The values of $E_{k,\rm{iso}}$ and $A_{\ast}$
adopted here can be found below. We adopt the convention of
$Q_{x}=Q/10^{x}$ in cgs units. Unless $n_0\ll0.1$ cm$^{-3}$,  the
jet has already decelerated to be non-relativistic ($\Gamma\sim2$)
while the fact that there was no jet break before
$t=4.1\times10^7$ s argues against the jet in an ISM medium. In
other words, the outflow of GRB 060729 is likely spherical if the
circum-burst medium is ISM. However, in this way the temporal
break at $T_{\rm break,4}$ can not be explained with the
hydrodynamic/geometry effect.

At such a late time, the jet may also enter the non-relativistic
phase in a stellar wind medium and the hydrodynamics is described
by the self-similar Sedov - von Neumann - Taylor solution. The
non-relativistic transition time\footnote{Zhang \& MacFadyen
(2009) suggested that the beginning time of the SNT self-similar
evolution is $\sim 5t_{\rm{NR}}$, where $t_{\rm{NR}}\sim
200E_{i,\rm{iso},54}A_{\ast}^{-1}$ years is the time when a GRB
jet with negligible sideways expansion becomes non-relativistic.}
is \citep{waxman04}
\begin{equation}
t_{\rm{SNT}} = 2.7 E_{\rm{jet},52}A_{\ast}^{-1} \;\rm{yr},
\end{equation}
where $E_{\rm{jet}}\simeq E_{k,\rm{iso}}\theta_{\rm{jet}}^2/4$ is
the beaming-corrected kinetic energy of the outflow. The
non-relativistic transition predicts a flattening in the light
curve \citep{huang03,zhangwq09}. The theoretical temporal slope is
$\alpha_5 = (7p-5)/6 \simeq 2.0 - 2.1$ for $p \sim 2.4 - 2.5$,
which is not inconsistent with the observations if the large error
bars of the observed slope are considered. However, the trend of
steepening after the break contradicts with the trend of
flattening when the jet becomes non-relativistic.

The jet is still inside the free wind bubble  $\sim 450$ days
after the burst in the observer's frame,  indicating that the
termination shock radius of the wind bubble is larger than $\sim
32$ pc. The size of a GRB progenitor star wind bubble depends on
the density and pressure of external interstellar medium, and the
mass loss history of the progenitor
\citep{garcia96,dai03,chevalier04}. The large size of the wind
bubble surrounding the progenitor of GRB 060729 suggests (1) a
fast and tenuous wind during the RSG phase prior to the Wolf-Rayet
phase, (2) the life time of the Wolf-Rayet stage is relatively
long, and (3) the density and pressure of the external
interstellar medium should be low. Therefore, from modeling the
afterglow, the progenitor of GRB 060729 is not likely to be
located in a giant molecular cloud or a star-burst environment.
Further deep optical observation of the GRB site and its host
galaxy may help to test this prediction.

\subsection{Spherical model}
We assume that the initial
shock-accelerated electrons at such a late time have a broken
power law distribution,
\begin{equation}
\frac{dN_e}{d\gamma_e}=N_{\gamma_e}\times\cases{\gamma_e^{-p_1}, &
$\gamma_m<\gamma_e<\gamma_b$, \cr
\gamma_b^{p_2-p_1}\gamma_e^{-p_2}, & $\gamma_b<\gamma_e$,}
\end{equation}
 where $p_1 = p \sim 2.4$ is the
low energy power law index and $p_2$ is the high energy power law
index, $\gamma_b$ is the break Lorentz factor of electrons. This
assumption is much more realistic than the single power law
assumption, especially for the late time when the shock Lorentz
factor decreases to the order of unity/the shock is no longer
ultra-relativistic (e.g., Hededal et al. 2004; Niemiec \&
Ostrowski 2006; Spitkovsky 2008). For simplicity, we assume
$R_b=\gamma_b/\gamma_m$ remains a constant in time (Li \&
Chevalier 2001). In this scenario, $t_c$ ($\nu_c = \nu_X$) is
equal to $1.3 \times 10^6$ s, $t_b$ ($\nu_b = \nu_X$, $\nu_b$ is
the typical synchrotron frequency of $\gamma_b$ electrons) is
equal to $4.1 \times 10^7$ s. The last steep decay segment can be
described with $\alpha_5 = (3p_2-1)/4$ and $\beta_5 = (p_2-1)/2$.
The high energy power index is therefore derived to be $p_2 =
6.53^{+2.73}_{-1.79}$, and the inferred spectral index is $\beta_5
= 2.77^{+1.37}_{-0.90}$. From $t_m \leq 10^5$ s and $t_b = 4.1
\times 10^7$ s, we constrain the parameter $R_b \geq
(4.1\times10^7/10^5)^{3/4} \sim 91$. The above $p_1 \sim 2.4$,
$p_2 \sim 6.5$ and $R_b \geq 91$ are quite similar to those
derived for GRBs 991208 and 000301C in Li \& Chevalier (2001).

The synchrotron emission from a spherical relativistic blast wave
can be described by (e.g., Chevalier \& Li 2000; Wu et al. 2005)
\begin{equation}
\nu_m = 3.7\times
10^{12}\epsilon_{e,-0.5}^2\epsilon_{B,-2.5}^{1/2}E_{k,\rm{iso},53}^{1/2}t_{5}^{-3/2}\;\rm{Hz},
\end{equation}
\begin{equation}
\nu_c = 3.4\times
10^{13}\epsilon_{e,-0.5}^{-1}\epsilon_{B,-2.5}^{-1/2}E_{k,\rm{iso},53}^{1/2}A_{\ast}^{-2}t_{5}^{1/2}\;\rm{Hz},
\end{equation}
\begin{equation}
F_{\nu,\rm{max}} =
111\epsilon_{B,-2.5}^{1/2}E_{k,\rm{iso},53}^{1/2}A_{\ast}t_{5}^{-1/2}\;\rm{mJy},
\end{equation}
where $\epsilon_e$, $\epsilon_B$ are the energy equipartition
fractions of electrons and magnetic field respectively, $A_{\ast}$
is the stellar wind parameter. We have considered the cooling
effect on $\nu_c$ by synchrotron-self-Compton scattering
processes. The crossing time $t_m$ in the optical band
($\nu_{\rm{opt}} \sim 10^{15}$ Hz) by $\nu_m$ no later than $t =
10^{5}$ s gives
\begin{equation}
\epsilon_{e,-0.5}^4\epsilon_{B,-2.5}E_{k,\rm{iso},53} \leq
7.3\times10^4,
\end{equation}
which is easily satisfied, while the crossing time $t_c$ in the
X-ray band ($h\nu_x \sim 2$ keV) by $\nu_c$ gives
\begin{equation}
\epsilon_{e,-0.5}\epsilon_{B,-2.5}^{1/2}E_{k,\rm{iso},53}^{-1/2}A_{\ast}^2
= 2.24\times10^{-4}.
\end{equation}
The flux density at frequency $\nu_c$, $F_{\nu_c}$, should be
equal to the observed 2 keV X-ray flux density of
$F_{2\rm{keV}}\sim3.5\times10^{-5}$ mJy at $t_c\sim10^{6}$ s,
which reads
\begin{equation}
\epsilon_{e,-0.5}^{-1.7}\epsilon_{B,-2.5}^{-0.7}E_{k,\rm{iso},53}^{2.4}A_{\ast}^{-5.2}
\sim 8.75\times10^{9}.
\end{equation}
The above afterglow model parameters can not be tightly
constrained due to the lack of more conditions/equations. However,
a large late-time isotropic kinetic energy of
$E_{k,\rm{iso}}\sim10^{54} - 10^{55}$ ergs and a small wind
parameter of $A_{\ast}\sim 0.1$ are favored if we assume
$\epsilon_{e}\sim0.1$ and $\epsilon_{B}\sim3\times10^{-3}$ as
inferred from fittings to other afterglows (e.g., Panaitescu \&
Kumar 2001; 2002). Within this parameter set,
$R_b=5.3\times10^4\epsilon_{e,-0.1}^{-1}\epsilon_{B,-2.5}^{-1/4}E_{k,\rm{iso},54}^{-1/4}$
is much higher than those derived from GRBs 991208 and 000301C by
Li \& Chevalier (2001).

\subsection{Energy Injection in the shallow decay phase}
A large $E_{k,\rm{iso}}$ ($\geq10^{54}$ ergs) at very late time
(after the plateau phase ends) can be understood if the early long
plateau phase detected both in X-ray and optical bands from a few
hundred seconds to $\sim 6\times10^4$ s since the trigger is
interpreted as the period of energy injection, as already
discussed in \citet{grupe07}. Possible scenarios for this enhanced
energy injection are refreshed shocks as suggested by
\citet{rees98} or the continuous energy injection by the strong
magnetic field of a newborn fast-rotating magnetar or black hole
as suggested by \citet{dai98} and \citet{zhang01}. In the latter
model, the temporal index $q$ of the injected luminosity
($L\propto t^{-q}$, $E_{\rm{inj}}\propto t^{1-q}$) can be derived
from the observed values of the temporal and spectral indices
during the plateau phase (e.g., Table 2 of Zhang et al. 2006).
From Table 5 of \citet{grupe07}, we can obtain
$q=2(\alpha_2-\beta_2+1)/(1+\beta_2)=-0.037\pm0.101$ with
$\alpha_2=0.14\pm0.02$ and $\beta_2=1.18\pm0.11$ for the X-ray
afterglow plateau assuming $\nu_m<\nu_c<\nu_X$. Such a $q$ value
is consistent with the theoretical expectation ($q=0$) of the
Poynting-flux dominated injection model. Therefore the isotropic
kinetic energy of the shock increases linearly with time,
$E_{k,\rm{iso}}\propto t$. The prompt isotropic gamma-ray energy
release in the $1$ keV - $10$ MeV band is
$E_{\gamma,\rm{iso}}=1.6\times10^{52}$ ergs (\citet{grupe07}).
Adopting a typical GRB efficiency of $10\% - 90\%$, the initial
isotropic kinetic energy remaining in the afterglow shock is about
the same order of $E_{\gamma,\rm{iso}}$, i.e.
$E_{k,\rm{iso},i}\sim 10^{52}$ ergs. After the energy injection is
finished, the shock energy $E_{k,\rm{iso}}$ is increased by a
factor of $\sim 100$ compared to its initial value, so
$E_{k,\rm{iso}}$ is the order of $\sim 10^{54}$ ergs. The lack of
a jet break up to $t=642$ days results a lower limit of the
half-opening angle of the GRB jet, i.e.,
\begin{equation}
\theta_{\rm j} \geq
15^{\circ}E_{k,\rm{iso},54}^{-1/4}A_{\ast,-1}^{1/4},
\end{equation}
which corresponds to a beaming-corrected jet (double-sided) energy
of
\begin{equation}
E_{\rm{jet}}\geq3.4\times10^{52}E_{k,\rm{iso},54}^{1/2}A_{\ast,-1}^{-1/2}\;\rm{ergs}.
\end{equation}

This makes it one of the most energetic jets ever seen and is the reason why we
were able to detect the X-ray afterglow still 642 days after the burst.
Note that the energy of a fastest-rotating magnetar is
$E_M=\frac{1}{2}I\Omega^2\leq2\times10^{52}I_{4 5}P_{\rm ms}^{-2}$
ergs. A massive and fast-rotating black hole as the central engine
is thus more likely to provide the required energy.

\section{Conclusion}

What makes the X-ray afterglow of GRB 060729 so remarkable is the
fact that it was still detected even almost two years after the
burst. This exceptional late-time detectability is related to
three things: a) with an initial 0.3 - 10.0 keV flux of almost $10^{-7}$
ergs s$^{-1}$ cm$^{-2}$ it was one of the brightest afterglows
ever detected by \swift\,,  b) its flat decay phase
\citep{nousek06,zhang06} extended out to about 60 ks after the
burst, and c) the decay slope after that break is about $\alpha$ =
1.3. Despite breaks at $T\sim1$ Ms and $T\sim1$ year the afterglow
was still detected by \chandra\ nearly two years after the burst.
Bursts like GRBs 060614, 061121, or even 080319B
\citep[][respectively]{mangano07, page07, racusin08} were even
brighter in X-rays at about 100 s after the burst than GRB 060729,
but their plateau phases are significantly shorter than that of
GRB 060729. They therefore faded more rapidly than GRB 060729 at late times.

Analysis and modeling of the X-ray afterglow of GRB 060729
show that this burst happened in a tenuous wind. This is
consistent with the collapsar picture of long GRBs. During the
early plateau phase, the energy in the external shock increased by
two orders of magnitude. A reanalysis of the \swift\, XRT light
curve together with the three detections by \chandra\, in 2007
reveal a temporal break at $\sim1.3$ Ms after the burst. The decay
slope steepened from $\alpha=1.32$ to $\alpha=1.61$ around this
break and the X-ray spectrum hardened in the meanwhile, indicating
this break is a cooling break (the cooling frequency of
synchrotron radiation crosses the X-ray band). There is another
light curve break at $\sim 1.3$ year after the burst tentatively
indicated by the last two \chandra\, detections. This break
coincides with a possible spectral softening, suggesting that the
break may be of spectral origin, though a hydrodynamic origin (jet
break) is also possible. If due to a jet break, then the implied
half-opening angle is $\theta_j \sim 14^{\circ}$. If due to a
spectral break, such a spectral softening could be the result of a
very steep power-law distribution of shock-accelerated electrons
responsible for the synchrotron radiation. In this case, with no
evidence for a jet break up to 642 days after the burst by
\chandra\, the jet half-opening angle must be
$\theta_j>15^{\circ}$ and the jet energy $E_j>3\times10^{52}$ erg.
Such a large jet energy implies that the central engine must be a
fast-rotating massive black hole, not a magnetar.

Our \chandra\ observations presented here have shown again how
important \chandra\ is for late-time observations of GRB X-ray
afterglows. \chandra\ has already been essential for the detection
and non-detection of jet breaks in the X-ray afterglows of the
short-duration GRBs 050724 and 051221A
\citep[][respectively]{grupe06,burrows06}.

\acknowledgments

We want to thank Sandy Patel and Chryssa Kouveliotou for providing
the data for the luminosities of \swift\ and pre-\swift\ bursts,
Hala Eid for fitting the late-time light curve with the
Beuermann et al. model, and Eric Feigelson for useful discussion about 
statistics.  
We thank the referee for his/her
suggestions/comments which have helped to improve our paper
significantly. 
We are extremely thankful to the whole \chandra\
team for successfully planning and performing the observations of
GRB 060729. XFW thanks Peter \meszaros, Kenji Toma, Derek Fox,
Antonino Cucchiara, and Yizhong Fan for their helpful discussion.
This work was also partially supported by NASA NNX 08AL40G (XFW),
National Natural Science Foundation of China (grants 10221001,
10403002, 10503012, 10621303, 10633040, and 10873002), and
National Basic Research Program of China (973 Program
2009CB824800) (XYW, XFW, and EWL). Swift is supported at Penn
State by NASA contract NAS5-00136.
 This research has been supported by
  SAO grants SV4-74018, A12 (D.G. and G.G.) and G08-9056 X (D.G.)

\clearpage



\clearpage

\begin{deluxetable}{lccrcc}
\tablecaption{\chandra\ Observation log of GRB 060729
\label{xray_log}}
\tablehead{
\colhead{ObsID} & \colhead{T-start\tablenotemark{1}} &
\colhead{T-stop\tablenotemark{1}} &
\colhead{$\rm T_{exp}$\tablenotemark{2}} & \colhead{CR\tablenotemark{3}}&
\colhead{$F_{\rm 0.3-10.0 keV}$\tablenotemark{4}}
}
\startdata
7567 & 2007-03-16 11:39 & 2007-03-16 19:57 & 27690 &  \\
8541 & 2007-03-17 13:02 & 2007-03-17 14:54 &  4701 & \rb{3.93$^{+1.28}_{-0.91}$} &
\rb{4.26$^{+1.39}_{-0.99}$} \\
\hline
7568 & 2007-05-16 09:30 & 2007-05-16 20:56 & 40268 & 2.67$^{+0.96}_{-0.66}$ & 2.90$^{+1.04}_{-0.72}$ \\
\hline
7569 & 2007-06-30 06:19 & 2007-06-30 23:48 & 60400 & 2.07$^{+0.72}_{-0.45}$ & 2.25$^{+0.78}_{-0.49}$ \\
\hline
9086 & 2007-12-26 01:02 & 2007-12-26 09:04 & 27395 & \\
9801 & 2007-12-28 19:23 & 2007-12-28 23:59 & 15068 & \\
9802 & 2007-12-29 18:57 & 2007-12-30 01:13 & 20239 & \rb{0.69$^{+0.35}_{-0.26}$} & \rb{0.75$^{+0.38}_{-0.28}$} \\
9803 & 2008-01-05 01:04 & 2008-01-05 04:23 &  9986 & \\
\hline
9087 & 2008-04-30 23:09 & 2008-05-01 08:34 & 32330 &  \\
9811 & 2008-05-01 22:45 & 2008-05-02 07:50 & 30775 & \\
9812 & 2008-05-03 15:51 & 2008-05-03 23:35 & 26382 & \rb{0.26$^{+0.18}_{-0.12}$} & \rb{0.28$^{+0.19}_{-0.13}$} \\
9813 & 2008-05-04 15:07 & 2008-05-04 23:22 & 27810 &
\enddata

\tablenotetext{1}{Start and End times are given in UT}
\tablenotetext{2}{Observing time given in s}
\tablenotetext{3}{Count rate in units of $10^{-4}$ ACIS-S counts
s$^{-1}$ in the 0.5 - 8.0 keV band} \tablenotetext{4}{Unabsorbed
0.3 - 10.0 keV flux in units of $10^{-15}$ ergs s$^{-1}$
cm$^{-2}$. The count rates were converted assuming
$N_H=1.34\times10^{21}$ cm$^{-2}$ and $\beta_{\rm x}$ = 0.89. }
\end{deluxetable}

\begin{deluxetable}{lcccccc}
\tablecaption{Fits to the late-time (T$>10^5$s) \swift\ XRT and
\chandra\ light curve of GRB 060729 \label{lc_fits}}
\tablehead{ \colhead{Model} & \colhead{$\alpha_{3}$}
&\colhead{$T_{\rm break,3}$\tablenotemark{1}} &
\colhead{$\alpha_{4}$} & \colhead{$T_{\rm
break,4}$\tablenotemark{1}} & \colhead{$\alpha_{5}$} &
\colhead{$\chi^2/\nu$} }

\startdata 1) powl fixed\tablenotemark{2} & 1.32 (fixed) & --- &
--- & --- & ---
&   897/46 \\
2) powl free\tablenotemark{3} & 1.45\plm0.01 & --- & --- & --- &
---
&   400/45 \\
3) bknpowl                     & 1.32$^{+0.03}_{-0.01}$ &
2.08$^{+0.42}_{-0.20}$ & 1.85$^{+0.10}_{-0.06}$ & --- & --- & 168/43 \\
4) powl\tablenotemark{4} & 1.46\plm0.04 & --- & --- & --- & --- &
160/41 \\
5) bknpowl\tablenotemark{4} & 1.32$^{+0.02}_{-0.05}$ &
1.23$^{+0.25}_{-0.40}$ &
1.70$^{+0.06}_{-0.07}$ & --- & --- & 65/39 \\
6) bknpowl $T<30$ Ms\tablenotemark{4} & 1.32$^{+0.02}_{-0.05}$ &
1.01$^{+0.35}_{-0.22}$ &  1.61$^{+0.10}_{-0.06}$ & ---
& --- & 60/37 \\
7) powl $T\geq1.2$ Ms\tablenotemark{4} & --- & --- & 1.68\plm0.08
&
---  & ---  & 12/15 \\
8) powl 1.2 Ms $<T<$ 35 Ms\tablenotemark{4} & --- & --- &
1.61$^{+0.07}_{-0.13}$ & --- & --- & 6/13 \\
9) bknpowl $T\geq1.2$ Ms\tablenotemark{4}$^,$\tablenotemark{5} &
--- &
--- & 1.61 (fixed)
& $41.3^{+4.2}_{-5.1}$ & $4.65^{+2.05}_{-1.34}$ & 6/14 \\
\enddata

\tablenotetext{1}{Break time $T_{\rm break}$ are given in units of
Ms} \tablenotetext{2}{Decay slope fixed to the value given in
\citet{grupe07}} \tablenotetext{3}{Decay slope parameter left free
to vary} \tablenotetext{4}{Excluding the two flares at 2 Ms and 5
Ms} \tablenotetext{5}{Error bars of $\alpha_5$ and $T_{\rm
break,2}$ are determined by keeping $\alpha_4$ fixed at the
best-fit value for model\,4 and assuming that $\chi^2-
\chi^2_{\rm{min}}$ follows the Gaussian distribution (see Fig.
\ref{contour}).}

\end{deluxetable}


\begin{thebibliography}{}
\bibitem[Arnaud(1996)]{arnaud96} Arnaud, K.~A., 1996, ASP
Conf.~Ser.~101: Astronomical Data Analysis Software and Systems V, 101, 17
\bibitem[Barthelmy(2005)]{barthelmy05} Barthelmy, S. D., 2005, Space Science
Reviews, 120, 143
\bibitem[Beuermann et al.(1999)]{beuermann99} Beuermann, K., et al., 1999, \aap, 352, L26
\bibitem[Burrows et al.(2005)]{burrows05} Burrows, D. N., et al., 2005, Space
Science Reviews, 120, 165
\bibitem[Burrows et al.(2006)]{burrows06} Burrows, D. N., et al., 2006, \apj, 653,
468
\bibitem[Burrows \& Racusin(2007)]{burrows07} Burrows, D. N., \& Racusin, J.,
2007, Il Nuovo Cimento B, Vol 121, 1273
\bibitem[Chevalier \& Li(2000)]{chevalier00} Chevalier, R. A., \& Li, Z.-Y., 2000,
\apj, 539, 195
\bibitem[Chevalier et al.(2004)]{chevalier04} Chevalier, R. A., Li, Z.-Y., \& Fransson, C., 2004,
\apj, 606, 369
\bibitem[Curran et al.(2008)]{curran08} Curran, P. A., van der Horst, A., \&
Wijers, R. A. M. J., 2008, \mnras, 386, 859
\bibitem[Dai \& Lu(1998)]{dai98} Dai, Z. G., \& Lu, T., 1998, \aap, 333, L87
\bibitem[Dai \& Wu(2003)]{dai03} Dai, Z. G., \& Wu, X. F., 2003, \apj, 591,
L21
\bibitem[Evans et al.(2009)]{evans09} Evans, P. A., et al., 2009, \mnras, 397,
1177
\bibitem[Frail et al.(2001)]{frail01} Frail, D. A., et al., 2001, \apj, 562, L55
\bibitem[Garcia-Segura et al.(1996)]{garcia96} Garcia-Segura, G., Langer, N., \& Mac Low, M-M, 1996,
\aap, 316, 133
\bibitem[Gehrels et al.(2004)]{gehrels04} Gehrels, N., et al., 2004, ApJ, 611,
1005
\bibitem[Granot \& Kumar(2003)]{granot2003} Granot, J., \& Kumar,
P., 2003, \apj, 591, 1086
\bibitem[Grupe et al.(2006)]{grupe06} Grupe, D., Burrows, D. N., Patel, S. K.,
Kouveliotou, C., Zhang, B., \meszaros, P., Wijers, R. A. M., \&
Gehrels, N., 2006, \apj, 653, 462
\bibitem[Grupe et al.(2007)]{grupe07} Grupe, D., et al., 2007, \apj, 662, 443
\bibitem[Hededal et al.(2004)]{hededal04} Hededal, C. B., Haugb{\o}lle, T.,
Frederiksen, J. T., \& Nordlund, \AA, 2004, \apj, 617, L107
\bibitem[Huang \& Cheng(2003)]{huang03} Huang, Y. F., \& Cheng,
K. S., 2003, MNRAS, 341, 263
\bibitem[Kumar \& Panaitescu(2000)]{kumar2000} Kumar,
P., \& Panaitescu, A., 2000, \apj, 541, L9
\bibitem[Kumar \& Granot(2003)]{kumar2003} Kumar,
P., \& Granot, J., 2003, \apj, 591, 1075
\bibitem[Li \& Chevalier(2001)]{li01} Li, Z.-Y., \& Chevalier, R. A., 2001, \apj,
551, 940
\bibitem[Liang et al.(2008)]{liang08} Liang, E.-W., Racusin, J. L., Zhang, B.,
Zhang, B.-B., \& Burrows, D. N., 2008, \apj, 675, 528
\bibitem[Mangano et al.(2007)]{mangano07} Mangano, V., et al., 2007, \aap, 470,
105
\bibitem[\meszaros(2006)]{meszaros06} \meszaros, P., 2006, Rep. Prog. Phys., 69, 2259
\bibitem[\meszaros et al.(1998)]{meszaros98} \meszaros, P., Rees, M. J., \& Wijers, R. A. M. J., 1998, \apj,
499, 301
\bibitem[Niemiec \& Ostroswski(2006)]{niemiec06} Niemiec, J., \& Ostroswski, M.,
2006, \apj, 641, 984
\bibitem[Nousek et al.(2006)]{nousek06} Nousek, J., Kouveliotou, C., Grupe, D.,
Page, K. L., et al., 2006, \apj, 642, 389
\bibitem[Page et al.(2007)]{page07} Page, K. L., et al., 2007, \apj, 663, 1125
\bibitem[Panaitescu \& Kumar(2001)]{panaitescu2001} Panaitescu, A., \& Kumar,
P., 2001, \apj, 560, L49
\bibitem[Panaitescu \& Kumar(2002)]{panaitescu2002} Panaitescu, A., \& Kumar,
P., 2002, \apj, 571, 779
\bibitem[Park et al.(2006)]{park06} Park, T., Kashyap, V. L., Siemiginowska, A., van
Dyk, D. A., Zezas, A., Heinke, C., \& Wargelin, B. J., 2006, \apj,
652, 610
\bibitem[Racusin et al.(2008)]{racusin08} Racusin, J. L., et al., 2008, Nature,
455, 183
\bibitem[Racusin et al.(2009)]{racusin09} Racusin, J. L., et al., 2009, \apj,
698, 43
\bibitem[Rees \& \meszaros(1998)]{rees98} Rees, M. J., \& \meszaros, P., 1998,
\apj, 496, L1
\bibitem[Rhoads(1999)]{rhoads99} Rhoads, J. E., 1999,
\apj, 525, 737
\bibitem[Roming et al.(2005)]{roming05} Roming, P. W. A., et al., 2005, Space
Science Reviews, 120, 95
\bibitem[Sari et al.(1998)]{sari98} Sari, R., Piran, T., \& Narayan, R., 1998,
\apj, 497, L17
\bibitem[Sari et al.(1999)]{sari99} Sari, R., Piran, T., \& Halpern, J. P., 1999,
\apj, 524, L43
\bibitem[Sato et al.(2007)]{sato07} Sato, G., et al., 2007, \apj, 657, 359
\bibitem[Spitkovsky (2008)]{spitkovsky08} Spitkovsky, A., 2008, \apj, 682, L5
\bibitem[Thoene et al.(2006)]{thoene06}
Thoene, C. C., et al. 2006, GCN Circ. 5373,
http://gcn.gsfc.nasa.gov/gcn3/ 5373.gcn3
\bibitem[Tiengo et al.(2003)]{tiengo03} Tiengo, A., Mereghetti, S., Ghisellini,
G., Rossi, E., Ghirlanda, G., \& Schartel, N., 2003, \aap, 409, 983
\bibitem[Tiengo et al.(2004)]{tiengo04} Tiengo, A., Mereghetti, S., Ghisellini,
G., Tavecchio, F., \& Ghirlanda, G., 2004, \aap, 423, 861
\bibitem[Troja et al.(2007)]{troja07} Troja, E., et al., 2007, \apj, 665, 599
\bibitem[Waxman(2004)]{waxman04} Waxman, E., 2004, \apj, 602, 886
\bibitem[Willingale et al.(2007)]{willingale07} Willingale, R., et al., 2007,
\apj, 662, 1093
\bibitem[Wu et al.(2005)]{wu05} Wu, X. F., Dai, Z. G., Huang, Y.
F., \& Lu, T., 2005, ApJ, 619, 968
\bibitem[Zhang \& \meszaros(2001)]{zhang01} Zhang, B., \& \meszaros, P, 2001,
\apj, 552, L35
\bibitem[Zhang \& \meszaros(2004)]{zhang04} Zhang, B., \& \meszaros, P., 2004, Int. Jour. Mod. Phys. A,
19, 2385
\bibitem[Zhang et al.(2006)]{zhang06} Zhang, B., et al., 2006, \apj, 642, 354
\bibitem[Zhang \& MacFadyen(2009)]{zhangwq09} Zhang, W. Q., \& MacFadyen, A., 2009,
\apj, 698, 1261
\end{thebibliography}
\end{document}